\documentclass[12pt]{article}
\usepackage{amsmath,amssymb,epsf,epsfig,amsthm,bm,graphicx}
\usepackage[dvips]{color}

\begin{document}

\title{ \flushright{\small CECS-PHY-08/06} \\
\center{Some considerations on the Mac Dowell-Mansouri action}}
\author{\textbf{Andr\'{e}s Anabal\'{o}n$^{1}$}\thanks{%
anabalon-at-cecs-dot-cl} \\
%EndAName
\textbf{\textit{$^{1}$}}{\small Centro de Estudios Cient\'{\i}ficos (CECS)
Casilla 1469 Valdivia, Chile}\\
{\small \textit{${}$}}\\
}
\maketitle

\begin{abstract}
A precise relation is established between the Stelle-West formulation of the
Mac Dowell-Mansouri approach to a gauge theory of gravity and the approach
based on a gauged Wess-Zumino-Witten term. In particular, it is shown that a
consistent truncation of the latter correspond to the former. A brief review
of the Lovelock-Chern-Simons motivation behind the gauged WZW ones is also
done.
\end{abstract}

\section{Introduction}

It has been a long struggle to describe the gravitational interaction in
terms of a gauge theory in the Yang-Mills sense. Up to now the results
appear to be very dissimilar in odd and even dimensions. While odd
dimensional Lovelock theories \cite{Lovelock:1971yv} can be used to
construct gauge theories of gravity, that, moreover, have a topological
interpretation using Chern-Simons (\textbf{CS}) forms \cite%
{Chamseddine:1989nu, Troncoso:1999pk}. Even dimensional Lovelock theories
appear to not be embeddable in topological structures.

However, it does exist the Mac Dowell-Mansouri \cite{MacDowell:1977jt} and
Chamseddine-West \cite{Chamseddine:1976bf} proposal (with the subsequent
Stelle-West improvement \cite{Stelle:1979aj}) to construct a gauge theory
for (super) gravity \textit{a la} Yang-Mills (in the sense that one of the
relevant objects in the construction is a Lie algebra valued connection).
This construction is elegant, somewhat reminiscent of having a topological
origin and have received some considerations through the years (see for
instance \cite{Wilczek:1998ea, Salgado:2003rf}).

On the other hand it has been pointed out that due to the natural connection
between Chern-Simons forms and gauged Wess-Zumino-Witten (\textbf{gWZW})
terms \cite{Anabalon:2006fj, Anabalon:2007dr} they should correspond to even
dimensional gauge theories of gravity. Since the above approach and the
Chamseddine-Mac Dowell-Mansouri-Stelle-West (\textbf{CMMSW}) one contains
similar ingredients, namely some $0$-form fields and a gauge connection, it
could be expected that they should share some similarity. In fact, as is
explicitly shown in this paper, they coincide when the non linear sigma
model of the gWZW is accordingly restricted.

The structure of this paper is as follow: first Lovelock theories and CS theories of gravity are briefly recasted, then the gWZW structure
is recalled and its properties are analyzed, the Unitary gauge is studied
and implemented and finally it is shown that a consistent restriction of the
gWZW theory exactly corresponds to the Stelle-West version of the CMMSW
theory.

\section{The Lovelock series}

A satisfactory description of nature is always accompanied by a reduced
number of assumptions. The main difficulty to reduce the number of
assumptions is that most of the times they are difficult to identify, and
even after identifying them it would be far from obvious how, in a sensible
way, relax them. Of course, these kind of considerations are relevant when
there is at hand a theory that has been proved to be physically sensible;
something that for the gravitational field, as described for the Einstein
field equations
\begin{equation}
R_{\mu v}-\frac{1}{2}g_{\mu v}R+\Lambda g_{\mu v}=8\pi GT_{\mu v},
\label{efi}
\end{equation}%
is supported by the experimental success associated with the description of
the primordial nucleosynthesis, the binary pulsar and of the solar system
tests \cite{Will:2005va}.

All this evidence, indeed suggests that the identification of the minimal
set of assumptions that implies (\ref{efi}), is a physically relevant
question. Luckily mathematicians think in uniqueness faster than physicists,
and Vermeil (1917), Weyl (1922) and Cartan (1922) showed (see \cite%
{Lovelock:1971yv} and references therein) that it is possible to single out
the LHS of (\ref{efi}), in every dimension, by asking

\begin{itemize}
\item A rank two, symmetric tensor

\item Covariant divergenceless.

\item Any derivative is at most second order and the tensor is linear in
them.
\end{itemize}

While the first two assumptions are motivated by what should appear at the
RHS of the Einstein tensor, and in fact are trivial if one begins with an
action principle instead of with field equations, the third is not so. As
was pointed out by Lovelock (1971) \cite{Lovelock:1971yv} it is possible to
relax linearity to quasi-linearity in the second derivatives (for a
discussion of quasi-linearity in this context see \cite{madore}).
Remarkably, this relaxation still implies that in four dimensions the only
possibility are the Einstein field equations, while, in higher dimensions,
gives rise to the Lovelock series.

A nice pattern that governs the Lovelock series is given by the
generalization of the relation between the Hilbert action and a two
dimensional topological invariant. The Hilbert action is a non-trivial
functional for the metric in all dimensions higher than two, while in two
dimension it becomes a boundary term known as the Euler density. The Euler
characteristic (the integral of the Euler density) is a number associated to
a family of manifolds that can be related by homotopies. It exist in all
dimensions, however it can be related with differentiable, geometrical
features of the manifold only if it is even dimensional, in which case is
given by the integral of

\begin{equation}
\frac{2\sqrt{\left\vert g\right\vert }}{4!4VOL(S^{4})}\delta _{\alpha \beta
\gamma \delta }^{\mu v\lambda \rho }R_{\text{ \ \ \ \ }\mu v}^{\alpha \gamma
}R_{\text{ \ \ \ }\lambda \rho }^{\lambda \delta }\text{, }\frac{2\sqrt{%
\left\vert g\right\vert }}{6!6VOL(S^{6})}\delta _{\alpha \beta \gamma \delta
\sigma \zeta }^{\mu v\lambda \rho \eta \tau }R_{\text{ \ \ \ \ }\mu
v}^{\alpha \gamma }R_{\text{ \ \ \ }\lambda \rho }^{\lambda \delta }R_{\text{
\ \ \ }\eta \tau }^{\sigma \zeta }\text{, ...}  \label{contrac}
\end{equation}%
\newline
where the quadratic term in the curvatures correspond to the four
dimensional case and the cubic to the six dimensional, the pattern in any
dimension follows from the above expression.

Any term of this series is: identically zero if the number of curvatures
that it contains, $p$, and the space-time dimension, $D$, is such that $2p>D$%
, does not contribute to the dynamics but are non trivial if $D=2p,$ and
gives rise to a term of the Lovelock series if $2p<D.$ This relation makes
people call the terms in Lovelock series the dimensional continuation of the
Euler density. Thus, the Lovelock series in dimension $D$ contains $\left[
\frac{D+1}{2}\right] $ terms, where $[\cdots]$ denotes the integer part. The
terms are the dimensionally continued Euler densities of all dimensions
below $D$, and the cosmological constant term.

Despite the condensed notation used in (\ref{contrac}), is possible to note
that the terms that can be added to the Lovelock Lagrangian increase its
complexity with the dimension. For instance the cubic one is proportional to
\cite{Mueller-Hoissen:1985mm}

\begin{align}
& 2R^{\alpha\beta\gamma\delta}R_{\gamma\delta\lambda v}R_{\text{ \ \ \ }%
\alpha\beta}^{\lambda v}+8R_{\text{ \ \ \ }\gamma\delta}^{\alpha\beta }R_{%
\text{ \ \ \ }\beta v}^{\gamma\lambda}R_{\text{ \ \ \ }\alpha\lambda
}^{\delta v}+24R^{\alpha\beta\gamma\delta}R_{\gamma\delta\beta v}R_{\alpha
}^{v} \\
& -3RR^{\alpha\beta\gamma\delta}R_{\alpha\beta\gamma\delta}+24R^{\alpha
\beta\gamma\delta}R_{\alpha\gamma}R_{\beta\delta}+16R^{\alpha\beta}R_{\beta%
\gamma}R_{\alpha}^{\gamma}-12RR^{\alpha\beta}R_{\alpha\beta}+R^{3}.  \notag
\end{align}

One equation is better than one thousand words, so the previous one is
enough to be convinced that a change in the notation is necessary to gain
insight in the Lovelock theory. To this end is necessary to introduce the
vielbein, $\bar{e}_{\mu }^{a}$, an isomorphism between the coordinate
tangent space and the non-coordinate one defined by the relation $\bar{e}%
_{\mu }^{a}\bar{e}_{v}^{b}\eta _{ab}=g_{\mu v}$ where $\eta _{ab}=diag\left(
-,+,..+\right) $.Using this isomorphism, the curvature two form
\begin{equation*}
R^{ab}\equiv \frac{1}{2}R_{\text{ \ \ \ }\mu v}^{ab}dx^{\mu }\wedge
dx^{v}\equiv \frac{1}{2}\bar{e}_{\alpha }^{a}\bar{e}_{\beta }^{b}R_{\text{ \
\ \ }\mu v}^{\alpha \beta }dx^{\mu }\wedge dx^{v},
\end{equation*}%
and the torsion two form
\begin{equation*}
T^{a}\equiv \frac{1}{2}T_{\mu v}^{a}dx^{\mu }\wedge dx^{v}\equiv \frac{1}{2}%
\bar{e}_{\gamma }^{a}T_{\mu v}^{\gamma }dx^{\mu }\wedge dx^{v}
\end{equation*}%
can be defined. They are related by means of the spin connection, $\omega
^{ab}\equiv \omega _{\mu }^{ab}dx^{\mu },$ through the identities

\begin{equation}
T^{a}\equiv d\bar{e}^{a}+\omega_{b}^{a}\wedge\bar{e}^{b}\equiv D\bar{e}%
^{a},\qquad R^{ab}\equiv d\omega^{ab}+\omega_{c}^{a}\wedge\omega^{cb},\qquad
DT^{a}=R^{ab}\wedge\bar{e}_{b}.
\end{equation}

Furthermore, using the convention that the wedge product ($\wedge $) is
assumed between forms, the Euler characteristic can be rewritten as the
integral of

\begin{equation}
\frac{2}{4!VOL(S^{4})}\varepsilon_{abcd}R^{ab}R^{cd}\text{, }\frac {2}{%
6!VOL(S^{6})}\varepsilon_{abcdef}R^{ab}R^{cd}R^{ef}\text{, ...}
\end{equation}
\newline
With this notation and the torsionless condition, $D\bar{e}^{a}=0$, the
Lovelock Lagrangians in four, five, six and seven dimensions can be written
as

\begin{align*}
\mathcal{L}_{4} & =\varepsilon_{abcd}\left( \alpha_{0}\bar{e}^{a}\bar {e}^{b}%
\bar{e}^{c}\bar{e}^{d}+\alpha_{1}\bar{e}^{a}\bar{e}^{b}R^{cd}\right) , \\
\mathcal{L}_{5} & =\varepsilon_{abcde}\left( \alpha_{0}\bar{e}^{a}\bar {e}%
^{b}\bar{e}^{c}\bar{e}^{d}\bar{e}^{e}+\alpha_{1}\bar{e}^{a}\bar{e}^{b}R^{cd}%
\bar{e}^{e}+\alpha_{2}\bar{e}^{a}R^{bc}R^{de}\right) \newline
, \\
\mathcal{L}_{6} & =\varepsilon_{abcdef}\left( \alpha_{0}\bar{e}^{a}\bar {e}%
^{b}\bar{e}^{c}\bar{e}^{d}\bar{e}^{e}+\alpha_{1}\bar{e}^{a}\bar{e}^{b}R^{cd}%
\bar{e}^{e}+\alpha_{2}\bar{e}^{a}R^{bc}R^{de}\right) \bar{e}^{f}, \\
\mathcal{L}_{7} & =\varepsilon_{abcdefg}\left( \alpha_{0}\bar{e}^{a}\bar {e}%
^{b}\bar{e}^{c}\bar{e}^{d}\bar{e}^{e}\bar{e}^{f}+\alpha_{1}\bar{e}^{a}\bar{e}%
^{b}R^{cd}\bar{e}^{e}\bar{e}^{f}+\alpha_{2}\bar{e}^{a}R^{bc}R^{de}\bar{e}%
^{f}+\alpha_{3}R^{ab}R^{cd}R^{ef}\right) \bar{e}^{g}.
\end{align*}

Where the $\alpha $ are dimensionful, arbitrary, coupling constants: $\alpha
_{0}$\ is proportional to the cosmological constant, $\alpha _{1}$\ is
related with the Newton constant while the remaining coupling constants are
related to the strength of its accompanying Lovelock term. This implies that
the most general Lovelock Lagrangian has $\left[ \frac{D+1}{2}\right] $
coupling constants, something that would ruin any possible interpretation of
it as a fundamental theory.

\subsection{Chern-Simons theories}

In the early eighties a related story, begun to evolve. A deep insight was
being obtained on background independent field theories; since all the
fundamental interactions needs the existence of a background metric to be
defined, background independence was mainly associated to the requirement
that the metric be a dynamical field. However, background independent field
theories can also be constructed beginning with no metric at all, to my
knowledge, this was pointed out to be the case by the first time with CS
theories \cite{Witten:1988hf}. The lack of the existence of any background
field implies a phase space implementation of diffeomorphism invariance,
that makes the CS theories similar to General Relativity (\textbf{GR}), and
it is in fact the case that, all the classical solutions of GR are contained
in a CS theory \cite{Achucarro:1987vz}, this highlighted the possibility of
show the exact solubility of the theory \cite{Witten:1988hc}. Notably
enough, nineteen years after these considerations, the relation between
gravity\ and CS\ gravity still is matter of research and apparently is far
from being completely understood \cite{Carlip:2005zn, Giacomini:2006dr,
Witten:2007kt}.

The CS\ formulation of 2+1 GR\ makes the theory explicitly power counting
renormalizable, this is because it can be reformulated in terms of a single
gauge connection,%
\begin{equation}
\mathcal{A}=\frac{1}{2}A_{\mu}^{AB}J_{AB}dx^{\mu}=\frac{1}{2}%
\omega^{ab}J_{ab}+\frac{\bar{e}^{a}}{l}J_{a3},  \label{con}
\end{equation}
\newline
where the vielbein is divided by a parameter with dimensions of length, $l,$
in order to make the one form $\frac{\bar{e}^{a}}{l}$ dimensionless. The
generators , $J_{AB},$ span the\ $SO(2,2)$ or $SO(3,1)$ algebras depending
if the cosmological constant is negative or positive. The Poincar\'{e} case
can be obtained by an In\"{o}n\"{u}-Wigner contraction of any of these cases.

Lets recall how the three dimensional Hilbert action can be rewritten as a
CS form (For a \ pedagogical review see \cite{Zanelli:2005sa})\footnote{$%
\langle ...\rangle$ stands for the invariant symmetric trace in the algebra,
$\langle J_{ab}J_{c3}\rangle=\varepsilon_{abc}.$}

\begin{align}
\kappa \int_{\Sigma }\left( R-2\Lambda \right) \sqrt{\left\vert g\right\vert
}d^{3}x& =\kappa \int_{\Sigma }\varepsilon _{abc}\bar{e}^{a}\left( R^{bc}\pm
\frac{1}{3l^{2}}\bar{e}^{b}\bar{e}^{c}\right)  \label{1} \\
& =\kappa l\int_{\Sigma }\varepsilon _{abc}e^{a}\left( R^{bc}\pm \frac{1}{3}%
e^{b}e^{c}\right)  \label{2} \\
& =\kappa l\int_{\Sigma }\left\langle \mathcal{A}d\mathcal{A}+\frac{2}{3}%
\mathcal{A}^{3}\right\rangle +\frac{\kappa l}{2}\int_{\Sigma }\varepsilon
_{abc}d\left( e^{a}\omega ^{bc}\right)  \label{3}
\end{align}%
where in (\ref{1}) the Palatini form of the Hilbert action is written in
terms $\bar{e}^{a}$, $\omega ^{ab}$ and $\Lambda =\mp \frac{1}{l^{2}}$. Note
that at this point the vielbein, $\bar{e}_{\mu }^{a}$, is an invertible
object that defines an isomorphism between the coordinate tangent space and
the non coordinate one. In (\ref{2}) the redefinition $\bar{e}=le$ was used.
In (\ref{3}) both objects, $\omega $ and $e,$ are put in the same foot,
making manifest the principal bundle structure of the theory.

The explicit power counting renormalizability motivated the search of a
higher dimensional realization of this structure, something done in \cite%
{Chamseddine:1989nu, Banados:1996hi}. CS forms exist in all odd dimensions,
thus further discomposing the connection in analogy with the three
dimensional case (\ref{con}) a particular class of gravities can be found,
one that contains higher powers in the curvature. It was latter realized
that this gravities can be supersymmetrized, but due to the lack of the
adequate superalgebras, the supersymmetrization of the CS gravities with
cosmological constant was stopped at dimension seven \cite%
{Chamseddine:1990gk}. Subsequent, exhaustive work, study most of the
possible supersymmetric versions of Chern-Simons gravities \cite%
{Banados:1996hi}.

Although the previous work was unrelated with the existence of Lovelock
gravity it gave a hint on how to solve a fundamental problem that it has,
namely the large number of, otherwise arbitrary, coupling constants present
in the theory. The relative values of the $\left[ \frac{D+1}{2}\right] $
coupling constants can be fixed by requiring that the local Lorentz
invariance, present in any Lovelock Lagrangian when written in terms of $e$
and $\omega $, enlarge to anti de Sitter, de Sitter or Poincar\'{e}
invariance. As was subsequently studied in \cite{Troncoso:1999pk} this
enlargement of the symmetry only occurs in odd dimensions, in which case the
Lovelock Lagrangian can be rewritten as a CS form.

As is discussed in \cite{Anabalon:2006fj, Anabalon:2007dr} CS theories and
gWZW forms are intriscally related, thus, they define our starting point to
construct a gauge theory of gravity in even dimensions.

\section{Four dimensional gWZW terms}

Seeking an effective lagrangian for pions it was suggested in \cite%
{Witten:1983tw} that a non-diagonally gauged version of the action principle%
\begin{align}
& S(h,\mathcal{A})=-\frac{\kappa }{10}\int_{M^{5}}\left\langle
h^{-1}dh(h^{-1}dh)^{2}(h^{-1}dh)^{2}\right\rangle  \notag \\
& +\kappa \int_{M^{4}}\left\langle dhh^{-1}\mathcal{A}\left( d\mathcal{A}+%
\frac{1}{2}\mathcal{A}^{2}\right) \right\rangle  \notag \\
& -\frac{\kappa }{2}\int_{M^{4}}\left\langle dhh^{-1}\mathcal{A}\left\{
(dhh^{-1})^{2}+\frac{1}{2}\left[ \mathcal{A}\,,dhh^{-1}\right] \right\}
\right\rangle  \notag \\
& -\kappa \int_{M^{4}}\left\langle \mathcal{A}\mathcal{A}^{h}\left( \mathcal{%
F}+\mathcal{F}^{h}-\frac{1}{2}\mathcal{A}^{2}-\frac{1}{2}(\mathcal{A}%
^{h})^{2}+\frac{1}{4}\left[ \mathcal{A},\mathcal{A}^{h}\right] \right)
\right\rangle \,,  \label{gwzw}
\end{align}%
where%
\begin{equation}
\mathcal{F}=d\mathcal{A}+\mathcal{AA},\mathcal{\qquad F}^{h}=h^{-1}\mathcal{F%
}h,\qquad \mathcal{A}^{h}=h^{-1}\mathcal{A}h+h^{-1}dh.
\end{equation}%
plus a kinetic term for the non-linear sigma model could represent the
searched action. In our perspective, the interest in the action (\ref{gwzw})
is that it is diffeomorphism invariant in the same sense that Chern-Simons
actions are; namely there is no necessity of a metric to define it. Thus,
they are perfectly adapted to describe gravitational theories.

In order to have a gravitational interpretation of (\ref{gwzw}) is necessary
to have a gauge group that contains the Lorentz group $so(3,1)$, and
furthermore in order to have a non-trivial WZW term one is obligated to
consider gauge groups that give rise to a trilinear invariant tensor. The
smaller algebras that satisfy the above conditions are $so(5,1)$, $so(4,2)$,
$so(3,3)$, with generators $J_{AB\text{ }}$and invariant tensor $%
\left\langle J_{AB}J_{CD}J_{EF}\right\rangle =\varepsilon _{ABCDEF}$. Along
the lines discussed in the introduction, $ISO(4,1)$, can also be considered.

Let%
%TCIMACRO{\U{b4}}%
%BeginExpansion
\'{}%
%EndExpansion
s recall some of the properties of the previous action: it is invariant
under the adjoint action of the gauge group, namely

\begin{equation}
\mathcal{A}\rightarrow g^{-1}\mathcal{A}g+g^{-1}dg,\qquad h\rightarrow
g^{-1}hg,
\end{equation}%
It can be noted that the action contains only even powers of $h,$\ and the
invariance

\begin{equation}
S(h,\mathcal{A})=S(-h,\mathcal{A})
\end{equation}%
follows from this fact.

A first suggestion that this theory could make sense is given by the relation

\begin{equation}
S(\mathcal{A}_{0},h_{0})=\kappa \sinh \theta _{0}\frac{3}{2}%
\int_{M^{4}}\varepsilon _{abcd}e^{a}e^{b}\left( R^{cd}+\mu e^{c}e^{d}\right)
\,.
\end{equation}

\begin{equation*}
\mathcal{A}_{0}=\frac{1}{2}\omega
^{ab}J_{ab}+e^{a}J_{a5}\;,\;\;h_{0}=e^{\theta _{0}J_{45}},\qquad \mu =\frac{1%
}{2}\left( 1-\cosh (\theta _{0})\right)
\end{equation*}%
where $\left( J_{ab},J_{a5}\right) $ span the $so(3,2)$ subalgebra of $%
so(4,2)$. However, there is no point to have a nice construction to later
mutilate it in order to obtain a desired result. Instead, if some condition
is going to be imposed on the field content of an action principle, it
should, at least, not modify the local symmetry present in the Lagrangian.

Parametrizing the non-linear sigma model as

\begin{equation}
h=\exp (\phi )=\exp (\frac{1}{2}J_{AB}\phi ^{AB}),
\end{equation}%
and using the Killing metric $Tr(J_{AB}J_{CD})=\eta _{AC}\eta _{BD}-\eta
_{BC}\eta _{AD}$\footnote{%
Here $\eta _{AB}=(-,+,+,+,+).$}, the following gauge invariant condition can
be imposed on the $\phi $ fields:

\begin{equation}
Tr(\phi \phi )=\frac{1}{4}\phi ^{AB}\phi ^{CD}\left( \eta _{AC}\eta
_{BD}-\eta _{BC}\eta _{AD}\right) =\frac{1}{2}\phi _{\text{ \ }}^{AC}\phi
_{AC}=m^{2}  \label{const}
\end{equation}%
where $m\,\ $is a constant. Indeed, restricting the field content to the
subspace defined by (\ref{const}) do not break the symmetry of (\ref{gwzw}),
and can be considered as a consistent restriction of it.

To further study the theory is neater to work in the Unitary gauge,
something elaborated in the next section.

\section{The field equations and the Unitary gauge.}

The field equations associated with the variation with respect to $h$ are
\begin{gather}
\kappa\int_{M^{4}}\Big\langle h^{-1}\delta h\Big\{\!\left( \mathcal{F}%
^{h}\right) ^{2}+\mathcal{F}^{2}+{}\!\mathcal{F}^{h}\mathcal{F}-\frac{3}{4}[%
\mathcal{A}^{h}-\mathcal{A},\mathcal{A}^{h}-\mathcal{A}]\ (\mathcal{F}^{h}+%
\mathcal{F})  \notag \\
+\frac{1}{8}[\mathcal{A}^{h}-\mathcal{A},\mathcal{A}^{h}-\mathcal{A}]^{2}+%
\frac{1}{2}(\mathcal{A}^{h}-\mathcal{A})[\mathcal{F}^{h}+\mathcal{F},%
\mathcal{A}^{h}-\mathcal{A}])\Big\}\Big\rangle,  \label{eomh}
\end{gather}
while those associated with the connection $\mathcal{A}$ are
\begin{equation}
\kappa\int_{M^{4}}\Big\langle\delta\mathcal{A}\left( (\mathcal{A}^{h}-%
\mathcal{A})\left( \mathcal{F}^{h}+2\mathcal{F}-\frac{1}{4}[\mathcal{A}^{h}-%
\mathcal{A},\mathcal{A}^{h}-\mathcal{A}]\right) \right) \Big\rangle%
-(h\leftrightarrow h^{-1})  \label{EOMA}
\end{equation}

\subsection{A relation between the field equations.}

The gauge invariance of the action allows to find off-shell identities
between the field equations. To see this, instead of the field variation of $%
(\mathcal{A},$ $h),$ is possible to begin with the fields $(\mathcal{A}^{g},$
$h^{g})=(g^{-1}\mathcal{A}g+g^{-1}dg,$ $g^{-1}hg)$ and consider the
variational derivatives of the action with respect to $\mathcal{A},$ $h$ and
$g$. This process gives the same field equations for the fields $(\mathcal{A}%
,$ $h)$ plus an identically satisfied extra contribution.

So, with the following variations

\begin{align}
\delta \left( \mathcal{A}^{g}\right) & =\delta g^{-1}\mathcal{A}g+g^{-1}%
\mathcal{A}\delta g+\delta g^{-1}dg+g^{-1}d\delta g+g^{-1}\delta \mathcal{A}g
\notag \\
& =g^{-1}\nabla \left( \delta gg^{-1}\right) g+g^{-1}\delta \mathcal{A}g, \\
\delta h& =g^{-1}\left[ h,\delta gg^{-1}\right] g+g^{-1}\delta hg,
\end{align}%
where $\nabla =d+\left[ \mathcal{A},\right] $, three extra terms are
obtained in the variational derivatives:

\begin{equation}
\int \left\langle g\delta g^{-1}h\mathcal{E}^{h}\mathcal{(A},h\mathcal{)}%
h^{-1}\right\rangle +\left\langle \delta gg^{-1}\mathcal{E}^{h}\mathcal{(A},h%
\mathcal{)}\right\rangle -\left\langle \delta gg^{-1}\nabla \mathcal{E}^{%
\mathcal{A}}\mathcal{(A},h\mathcal{)}\right\rangle .
\end{equation}%
where $\mathcal{E}^{h}\mathcal{(A},h\mathcal{)}$ are the field equations of $%
h$ and $\mathcal{E}^{\mathcal{A}}\mathcal{(A},h\mathcal{)}$ of\ $\mathcal{A}%
. $ Gauge invariance implies that this relation is identically satisfied.
Thus, it follows that%
\begin{equation}
\mathcal{E}^{h}\mathcal{(A},h\mathcal{)}-h\mathcal{E}^{h}\mathcal{(A},h%
\mathcal{)}h^{-1}=\nabla \mathcal{E}^{\mathcal{A}}\mathcal{(A},h\mathcal{)}
\label{OF}
\end{equation}%
which means that the consistence condition, $\nabla \mathcal{E}^{\mathcal{A}}%
\mathcal{(A},h\mathcal{)},$ is trivially satisfied when the field equations
for $h$, $\mathcal{E}^{h}\mathcal{(A},h\mathcal{)}$, holds. The last
identity can be checked explicitly replacing the field equations at both
sides of it.

\subsection{A decomposition for h}

An arbitrary element of a semisimple Lie algebra can be written as the
adjoint action of the lie algebra on a Cartan subalgebra. So, the following
local decomposition follows,%
\begin{equation}
h=p^{-1}ap.  \label{DESC}
\end{equation}
Using (\ref{DESC})\ the task of solving the field equations simplifies:

\begin{align}
\left\langle h^{-1}\delta h\mathcal{E}^{h}\mathcal{(A},h\mathcal{)}%
\right\rangle & =\left\langle h^{-1}\left( \delta p^{-1}ap+p^{-1}\delta
ap+p^{-1}a\delta p\right) \mathcal{E}^{h}\mathcal{(A},h\mathcal{)}%
\right\rangle \\
& =\left\langle \left( h^{-1}\delta p^{-1}ph+p^{-1}a^{-1}\delta
ap+p^{-1}\delta p\right) \mathcal{E}^{h}\mathcal{(A},h\mathcal{)}%
\right\rangle \\
& =\left\langle p^{-1}\delta p\left( -h\mathcal{E}^{h}\mathcal{(A},h\mathcal{%
)}h^{-1}+\mathcal{E}^{h}\mathcal{(A},h\mathcal{)}\right) \right\rangle
\label{ma} \\
& +\left\langle a^{-1}\delta ap\mathcal{E}^{h}\mathcal{(A},h\mathcal{)}%
p^{-1}\right\rangle  \notag \\
& =\left\langle a^{-1}\delta a\mathcal{E}^{h}\mathcal{(B},a\mathcal{)}%
\right\rangle +\left\langle p^{-1}\delta p\nabla \mathcal{E}^{\mathcal{A}}%
\mathcal{(A},h\mathcal{)}\right\rangle  \label{gic} \\
& =\left\langle a^{-1}\delta a\mathcal{E}^{h}\mathcal{(B},a\mathcal{)}%
\right\rangle +\left\langle \delta pp^{-1}\bar{\nabla}\mathcal{E}^{\mathcal{A%
}}\mathcal{(B},a\mathcal{)}\right\rangle ,  \label{rew}
\end{align}%
\newline
where $\mathcal{B}\equiv p\mathcal{A}p^{-1}+pdp^{-1},$\ $\bar{\nabla}=d+%
\left[ \mathcal{B},\right] $ and (\ref{OF})$\ $\ was used to pass from (\ref%
{ma}) to (\ref{gic}). On the other hand, the field equations for $\mathcal{A}
$ can be rewritten as

\begin{equation}
\left\langle J_{AB}\mathcal{E}^{\mathcal{A}}\mathcal{(B},a\mathcal{)}%
\right\rangle =0.  \label{feq}
\end{equation}%
So, $p$\ is in fact pure gauge, since it is not determined by any field
equation. Thus, one is left with the milder task of solving the field
equations in the so-called unitary gauge

\begin{equation}
\left\langle J_{AB}\mathcal{E}^{\mathcal{A}}\mathcal{(B},a\mathcal{)}%
\right\rangle =0\qquad \left\langle C_{AB}\mathcal{E}^{h}\mathcal{(B},a%
\mathcal{)}\right\rangle =0.  \label{feq un}
\end{equation}%
where $C_{AB}$ are generators along a Cartan subalgebra.

Note that the above deduction considered that all the fields of the
non-linear sigma model where independently varied, while if the condition (%
\ref{const}) is taken into account one has that, in the unitary gauge

\begin{equation}
-\phi ^{01}\delta \phi ^{01}+\phi ^{23}\delta \phi ^{23}-\phi ^{45}\delta
\phi ^{45}=0
\end{equation}%
{}

\begin{equation}
\Longrightarrow \delta \phi ^{45}=\frac{-\phi ^{01}\delta \phi ^{01}+\phi
^{23}\delta \phi ^{23}}{\phi ^{45}}
\end{equation}

\begin{equation}
\Longrightarrow a^{-1}\delta a=\delta \phi ^{01}\left( J_{01}-\frac{\phi
^{01}}{\phi ^{45}}J_{45}\right) +\delta \phi ^{23}\left( J_{23}+\frac{\phi
^{23}}{\phi ^{45}}J_{45}\right)  \label{var}
\end{equation}

Now, the main problem to obtain Einstein gravity from the gWZW term is that
equations quadratic in the curvature arise when the field equations
associated to the non-linear sigma model are taken in account \cite%
{Anabalon:2006fj, Anabalon:2007dr}. Thus, in configurations of constant $%
\phi $, the system become overconstrained and, for instance, the unique
spherically symmetric solution is flat space \cite{Giribet:2006ec}. This
quadratic constraint is proportional to a four form times $\varepsilon
_{abcd}$, so it appears from the field equation of the form $\left\langle
J_{45}\mathcal{E}^{h}\mathcal{(B},a\mathcal{)}\right\rangle $. The
restricted set of variations defined by (\ref{var}) imply that it disappear
when the $\phi $ field take some trivial values.

The simplest case to examine the above construction explicitly is when $
ISO(4,1)$ is set as the gauge group, it is the subject of the next section.

\section{The ISO(4,1) case}

It is convenient to consider the $iso(4,1)$ case because it contains an
abelian, invariant, subalgebra. It allows to restrict the non-linear sigma
model to take its values on this subalgebra without affecting the local
symmetry of the action. Decomposing the $iso(4,1)$ algebra in its $so(3,1)$
irreducible parts the generators reads $(J_{ab},P_{c},T_{c},W)$, where $%
(J_{ab},P_{c})$ span the $iso(3,1)$ subalgebra and $(J_{ab},T_{c})$ span the
$so(4,1)$ subalgebra. The commutation relations are

\begin{eqnarray}
\left[ J_{ab},J_{cd}\right]  &=&-J_{ac}\eta _{bd}+J_{bc}\eta
_{ad}-J_{bd}\eta _{ac}+J_{ad}\eta _{bc}, \\
\left[ J_{ab},T_{c}\right]  &=&-T_{b}\eta _{ac}+T_{a}\eta _{bc},\qquad \left[
J_{ab},P_{c}\right] =-P_{b}\eta _{ac}+P_{a}\eta _{bc}, \\
\left[ T_{a},P_{c}\right]  &=&-W\eta _{ac},\qquad \left[ T_{a},W\right]
=P_{a},\qquad \left[ T_{a},T_{b}\right] =-J_{ab}.
\end{eqnarray}%
\begin{equation}
a=0,...,3\;\;\eta _{ab}=\left( -,+,+,+,\right)
\end{equation}%
and, correspondingly, the connection is written as
\begin{equation}
\mathcal{A}=\frac{1}{2}\omega ^{ab}J_{ab}+c^{a}P_{a}+b^{a}T_{a}+\Phi W\;,
\end{equation}%
while the curvature reads
\begin{equation}
\mathcal{F}=\frac{1}{2}\left( R^{ab}-b^{a}b^{b}\right) J_{ab}+\left(
db^{a}+\omega ^{ac}b_{c}\right) T_{a}+\left( dc^{a}+\omega
^{ab}c_{b}+b^{a}\Phi \right) P_{a}+\left( d\Phi -b^{a}c_{a}\right) W.
\end{equation}%
The simplest thing that one can do is to consider that the non-linear sigma
takes its values along the generators $(P,W)$:%
\begin{equation}
h=\exp (z^{A}P_{A}),\qquad P_{A}=(P_{a},W),
\end{equation}%
In this way the gWZW action takes the simple form

\begin{equation}
S(h,\mathcal{A})=3\kappa \int_{M^{4}}z^{A}\varepsilon _{ABCDE}\Omega
^{BC}\Omega ^{DE}\,,  \label{cha}
\end{equation}%
{}

\begin{equation}
\Omega =\frac{1}{2}\Omega ^{AB}J_{AB}=\frac{1}{2}\left(
R^{ab}-b^{a}b^{b}\right) J_{ab}+\left( db^{a}+\omega ^{ac}b_{c}\right) T_{a},
\end{equation}%
which after imposing the gauge invariant constraint $z^{A}z_{A}=m^{2}$,
gives rise to standard Einstein gravity. In the above action part of the
original $ISO(4,1)$ symmetry is realized in a trivial way and the remanent
symmetry is just $SO(4,1)$.

Thus, we have exactly reproduced de CMMSW gauge theory of gravity. Too much
exactly; the main two drawbacks of the this theory are still present \cite%
{Wilczek:1998ea}. That is the necessity to impose the gauge invariant
constraint (\ref{const}) by hand and the lack of a good reason to consider a
sector of the gauge connection to be invertible (the vielbein).
Interestingly enough the second of these issues is solved by a relation that
looks exactly like a term of (\ref{eomh}) (see equation 14 in \cite%
{Wilczek:1998ea}), something that would deserve further consideration.

\textbf{Acknowledgments}

The author would like to thank Ricardo Troncoso for encouragement to the
realization of this work. This work is supported by the grant No. 3080024
from FONDECYT (Chile). The Centro de Estudios Cient\'{\i}ficos (CECS) is
funded by the Chilean Government through the Millennium Science Initiative
and the Centers of Excellence Base Financing Program of Conicyt. CECS is
also supported by a group of private companies which at present includes
Antofagasta Minerals, Arauco, Empresas CMPC, Indura, Naviera Ultragas and
Telef\'{o}nica del Sur.


\begin{thebibliography}{99}
\bibitem{Lovelock:1971yv} D.~Lovelock,
%``The Einstein tensor and its generalizations,''
J.\ Math.\ Phys.\ \textbf{12} (1971) 498. %%CITATION = JMAPA,12,498;%%

\bibitem{Chamseddine:1989nu} A.~H.~Chamseddine,
%``TOPOLOGICAL GAUGE THEORY OF GRAVITY IN FIVE-DIMENSIONS AND ALL ODD  DIMENSIONS,''
Phys.\ Lett.\ B \textbf{233} (1989) 291. %%CITATION = PHLTA,B233,291;%%

\bibitem{Troncoso:1999pk} R.~Troncoso and J.~Zanelli,
%``Higher dimensional gravity and local AdS symmetry,''
Class.\ Quant.\ Grav.\ \textbf{17} (2000) 4451 [arXiv:hep-th/9907109].
%%CITATION = CQGRD,17,4451;%%

\bibitem{MacDowell:1977jt} S.~W.~MacDowell and F.~Mansouri,
%``Unified Geometric Theory Of Gravity And Supergravity,''
Phys.\ Rev.\ Lett.\ \textbf{38} (1977) 739 [Erratum-ibid.\ \textbf{38}
(1977) 1376]. %%CITATION = PRLTA,38,739;%%

\bibitem{Chamseddine:1976bf} A.~H.~Chamseddine and P.~C.~West,
%``Supergravity As A Gauge Theory Of Supersymmetry,''
Nucl.\ Phys.\ B \textbf{129} (1977) 39. %%CITATION = NUPHA,B129,39;%%

\bibitem{Stelle:1979aj} K.~S.~Stelle and P.~C.~West,
%``Spontaneously Broken De Sitter Symmetry And The Gravitational Holonomy
%Group,''
Phys.\ Rev.\ D \textbf{21} (1980) 1466. %%CITATION = PHRVA,D21,1466;%%

\bibitem{Wilczek:1998ea} F.~Wilczek,
%``Riemann-Einstein structure from volume and gauge symmetry,''
Phys.\ Rev.\ Lett.\ \textbf{80} (1998) 4851 [arXiv:hep-th/9801184].
%%CITATION = PRLTA,80,4851;%%

\bibitem{Salgado:2003rf} P.~Salgado, F.~Izaurieta and E.~Rodriguez,
%``Higher dimensional gravity invariant under the AdS group,''
Phys.\ Lett.\ B \textbf{574} (2003) 283 [arXiv:hep-th/0305180].
%%CITATION = PHLTA,B574,283;%%  
  H.~Garcia-Compean, J.~A.~Nieto, O.~Obregon and C.~Ramirez,
  %``Dual description of supergravity MacDowell-Mansouri theory,''
  Phys.\ Rev.\  D {\bf 59} (1999) 124003
  [arXiv:hep-th/9812175].
  %%CITATION = PHRVA,D59,124003;%%
  J.~A.~Nieto, J.~Socorro and O.~Obregon,
  %``Gauge Theory Of Supergravity Based Only On A Selfdual Spin Connection,''
  Phys.\ Rev.\ Lett.\  {\bf 76} (1996) 3482.
  %%CITATION = PRLTA,76,3482;%%
\bibitem{Anabalon:2006fj} A.~Anabal\'{o}n, S.~Willison and J.~Zanelli,
%``General relativity from a gauged WZW term,''
Phys.\ Rev.\ D \textbf{75} (2007) 024009 [arXiv:hep-th/0610136].

\bibitem{Anabalon:2007dr} A.~Anabalon, S.~Willison and J.~Zanelli,
%``The Universe as a topological defect,''
Phys.\ Rev.\ D \textbf{77} (2008) 044019 [arXiv:hep-th/0702192].
%%CITATION = PHRVA,D77,044019;%%

\bibitem{Will:2005va} C.~M.~Will,
%``The confrontation between general relativity and experiment,''
arXiv:gr-qc/0510072. %%CITATION = GR-QC/0510072;%%

\bibitem{madore} N. Deruelle and J. Madore, arXiv:gr-qc/0305004.
%%%CITATION = GR-QC/0305004;%%

\bibitem{Mueller-Hoissen:1985mm} F.~Mueller-Hoissen,
%``Spontaneous Compactification With Quadratic And Cubic Curvature Terms,''
Phys.\ Lett.\ B \textbf{163} (1985) 106. %%CITATION = PHLTA,B163,106;%%

\bibitem{Witten:1988hf} E.~Witten,
%``Quantum field theory and the Jones polynomial,''
Commun.\ Math.\ Phys.\ \textbf{121} (1989) 351.
%%CITATION = CMPHA,121,351;%%

\bibitem{Achucarro:1987vz} A.~Ach\'ucarro and P.~K.~Townsend,
%``A CHERN-SIMONS ACTION FOR THREE-DIMENSIONAL ANTI-DE SITTER SUPERGRAVITY
%THEORIES,''
Phys.\ Lett.\ B \textbf{180} (1986) 89. %%CITATION = PHLTA,B180,89;%%

\bibitem{Witten:1988hc} E.~Witten,
%``(2+1)-Dimensional Gravity as an Exactly Soluble System,''
Nucl.\ Phys.\ B \textbf{311}, 46 (1988). %%CITATION = NUPHA,B311,46;%%

\bibitem{Carlip:2005zn} S.~Carlip,
%``Conformal field theory, (2+1)-dimensional gravity, and the BTZ black
%hole,''
Class.\ Quant.\ Grav.\ \textbf{22} (2005) R85 [arXiv:gr-qc/0503022].
%%CITATION = CQGRD,22,R85;%%

\bibitem{Giacomini:2006dr} A.~Giacomini, R.~Troncoso and S.~Willison,
%``Three-dimensional supergravity reloaded,''
Class.\ Quant.\ Grav.\ \textbf{24}, 2845 (2007) [arXiv:hep-th/0610077].
%%CITATION = CQGRD,24,2845;%%

\bibitem{Witten:2007kt} E.~Witten,
%``Three-Dimensional Gravity Revisited,''
arXiv:0706.3359 [hep-th]. %%CITATION = ARXIV:0706.3359;%%

\bibitem{Zanelli:2005sa} J.~Zanelli, \textquotedblleft Lecture notes on
Chern-Simons (super-)gravities,\textquotedblright\ [arXiv:hep-th/0502193].
%%CITATION = HEP-TH 0502193;%%

\bibitem{Chamseddine:1990gk} A.~H.~Chamseddine,
%``Topological gravity and supergravity in various dimensions,''
Nucl.\ Phys.\ B \textbf{346} (1990) 213. %%CITATION = NUPHA,B346,213;%%

\bibitem{Banados:1996hi} M.~Banados, R.~Troncoso and J.~Zanelli,
%``Higher dimensional Chern-Simons supergravity,''
Phys.\ Rev.\ D \textbf{54} (1996) 2605 [arXiv:gr-qc/9601003].%
%%CITATION = PHRVA,D54,2605;%%
R.~Troncoso and J.~Zanelli,
%``New gauge supergravity in seven and eleven dimensions,''
Phys.\ Rev.\ D \textbf{58} (1998) 101703 [arXiv:hep-th/9710180].
%%CITATION = PHRVA,D58,101703;%%
R.~Troncoso and J.~Zanelli,
%``Gauge supergravities for all odd dimensions,''
Int.\ J.\ Theor.\ Phys.\ \textbf{38} (1999) 1181 [arXiv:hep-th/9807029].
%%CITATION = IJTPB,38,1181;%%
R.~Troncoso and J.~Zanelli,
%``Chern-Simons supergravities with off-shell local superalgebras,''
arXiv:hep-th/9902003. %%CITATION = HEP-TH/9902003;%%
M.~Hassaine, R.~Olea and R.~Troncoso,
%``New N = 2 supergravity with local Poincare invariance in nine dimensions,''
Phys.\ Lett.\ B \textbf{599} (2004) 111 [arXiv:hep-th/0210116].
%%CITATION = PHLTA,B599,111;%%
M.~Hassaine, R.~Troncoso and J.~Zanelli,
%``Eleven-dimensional supergravity as a gauge theory for the M-algebra,''
Phys.\ Lett.\ B \textbf{596} (2004) 132 [arXiv:hep-th/0306258].
%%CITATION = PHLTA,B596,132;%%
M.~Hassaine, R.~Troncoso and J.~Zanelli,
%``11D Supergravity as a gauge theory for the M-algebra,''
PoS \textbf{WC2004} (2005) 006 [arXiv:hep-th/0503220].
%%CITATION = POSCI,WC2004,006;%%
M.~Hassaine and M.~Romo,
%``Local supersymmetric extensions of the Poincare and AdS invariant
%gravity,''
arXiv:0804.4805 [hep-th]. %%CITATION = ARXIV:0804.4805;%%

\bibitem{Witten:1983tw} E.~Witten, %``Global Aspects Of Current Algebra,''
Nucl.\ Phys.\ B \textbf{223}, 422 (1983). %%CITATION = NUPHA,B223,422;%%

\bibitem{Giribet:2006ec} G.~Giribet, J.~Oliva and R.~Troncoso,
%``Simple compactifications and black p-branes in Gauss-Bonnet and Lovelock
%theories,''
JHEP \textbf{0605}, 007 (2006) [arXiv:hep-th/0603177].
%%CITATION = JHEPA,0605,007;%%
\end{thebibliography}
\end{document}